\documentclass[useAMS,usenatbib]{mn2e} 
\usepackage{aas_macros}
\usepackage{graphics}
\usepackage[pdftex]{graphicx}
\usepackage{epstopdf}
\usepackage{epsfig}  
\usepackage{natbib} 
\usepackage{float}
\usepackage{amsmath}
\usepackage{times}
\usepackage[varg]{txfonts}
\usepackage{verbatim} % need that for multi-line comments
\usepackage{multirow,bigdelim} % to do the large brackets in the mock table
\usepackage{color}
\usepackage{vmargin}
\setmarginsrb{1.2cm}{1cm}{1.2cm}{1cm}{1cm}{1cm}{1cm}{1cm}

\newcommand{\hmpc}{{\,\rm h^{-1}Mpc}}

  \makeatletter
    \renewcommand{\paragraph}{\@startsection{paragraph}{4}{\z@}%
      {-3.25ex\@plus -1ex \@minus -.2ex}%
      {1.5ex \@plus .2ex}%
      {\normalfont\small\centering}}
     
    \renewcommand{\subparagraph}{\@startsection{subparagraph}{5}{\z@}%
      {-3.25ex\@plus -1ex \@minus -.2ex}%
      {1.5ex \@plus .2ex}%
      {\normalfont\small\centering}}
    \makeatother

\newcommand{\gadgettwo}{{\sc Gadget-2}}

\newcommand{\kms}{{ km s$^{-1}$}}
\newcommand{\hMpc}{{ h$^{-1}$ Mpc}}

\title[Constrained Simulations]{Simulations of the Local Universe Constrained by Observational Peculiar Velocities}
\author[Sorce et al.]
{
Jenny G. Sorce$^{1,2}$\thanks{E-mail: \texttt{j.sorce@ipnl.in2p3.fr}}, 
H\'el\`ene M. Courtois$^{1}$,
Stefan Gottl\"{o}ber$^2$,
Yehuda Hoffman$^3$,
R. Brent Tully$^{4}$\\
$^{1}$Universit\'e Lyon 1, CNRS/IN2P3, Institut de Physique Nucl\'eaire, Lyon, France\\
$^2$Leibniz-Institut f\"{u}r Astrophysik, Potsdam, Germany\\
$^3$Racah Institute of Physics, Hebrew University, Jerusalem, Israel\\
$^4$Institute for Astronomy, University of Hawaii, 2680 Woodlawn Drive, HI 96822, USA\\
}

\begin{document}

\date{}

\pagerange{\pageref{firstpage}--\pageref{lastpage}} \pubyear{2014}

\maketitle

\label{firstpage}

%typical abstract:context, aims, methods, results and conclusions + keywords
\begin{abstract}
\indent Peculiar velocities, obtained from direct distance measurements, are data of choice to achieve constrained simulations of the Local Universe reliable down to a scale of a few Megaparsecs. Unlike redshift surveys, peculiar velocities are direct tracers of the underlying gravitational field as they trace both baryonic and dark matter.  This paper presents the first attempt to use solely observational peculiar velocities to constrain cosmological simulations of the nearby universe.
In order to set up Initial Conditions, a Reverse Zel'dovich Approximation (RZA) is used to displace constraints from their positions at z=0 to their precursors' locations at higher redshifts. An additional new feature replaces original observed radial peculiar velocity vectors by their full 3D reconstructions provided by the Wiener-Filter (WF) estimator. Subsequently,  the Constrained Realization of Gaussian fields technique (CR) is applied to build various realizations of the Initial Conditions.
  The WF/RZA/CR method is first tested on realistic mock catalogs built from a reference simulation similar to the Local Universe. These mocks include errors on peculiar velocities, on data-point positions and a large continuous zone devoid of data in order to mimic galactic extinction. % The obtained constrained simulations allow to test the cosmic variance in a small volume. 
 Large scale structures are recovered with a typical accuracy of  5 h$^{-1}$ Megaparsecs in position, the best realizations reaching a 2-3 \hMpc\ precision, the limit imposed by the RZA linear theory.
  Then, the method is applied to the first observational radial peculiar velocity catalog of the project Cosmicflows. This paper is a proof of concept that the WF/RZA/CR method can be applied to observational peculiar velocities to successfully build constrained Initial Conditions.
  
\end{abstract}

%%%%%%%%%%%%%%%%%%%%%%%%%%%%%%%%%%%%%%%%%%%%%%%%%%%%%%%%%%%%%
%%%%%%%%%%%%%%%%%%%%%%%%%%%%%%%%%%%%%%%%%%%%%%%%%%%%%%%%%%%%%
%INTRODUCTION%%%%%%%%%%%%%%%%%%%%%%%%%%%%%%%%%%%%%%%%%%%%%%%%%%%%
%%%%%%%%%%%%%%%%%%%%%%%%%%%%%%%%%%%%%%%%%%%%%%%%%%%%%%%%%%%%%
%%%%%%%%%%%%%%%%%%%%%%%%%%%%%%%%%%%%%%%%%%%%%%%%%%%%%%%%%%%%%

\section{Introduction}

Under the assumption of a dark-matter only Universe, even the simplest problem of the emergence of structures defies a proper and detailed analytical treatment. As a result, the study of the formation of the Large Scale Structure of the Universe relies heavily on numerical simulations. Large scale dark matter simulations \citep[e.g.,][]{Klypin2011,DeusSimulation2012,Prada2012,AnguloXXL2012,Jubilee2013} constitute the backbone of the study of structure formation in the Universe. 

The standard model of cosmology asserts that  the primordial fluctuations are constituted by a Gaussian random field whose statistical properties are determined by its power spectrum. An accurate determination of the power spectrum is enabled by the cosmological parameters. These latter are obtained from observations of the fluctuations in the cosmic microwave background radiation combined with Baryonic Acoustic Oscillations and supernova measurements \citep{2011ApJS..192...18K,2011ApJS..192...16L,2013arXiv1303.5076P}. 
Standard cosmological computations use Initial Conditions (ICs) drawn from random realizations of the primordial perturbation field. 

By contrast, \emph {constrained simulations} stem from ICs obeying a set of observational constraints in addition to the random component. They provide a different approach to cosmological simulations to better approximate the observed nearby Universe. Constraints can either be peculiar velocities or galaxy distributions. The first constrained ICs were produced by \citet{1993ApJ...415L...5G}, using the Mark III catalog of peculiar velocities \citep{1996yCat.7198....0W}. These ICs were then used to perform the first constrained simulation of the nearby universe by \citet{1996ApJ...458..419K}. The CLUES (Constrained Local UniversE Simulations) project has been running a variety of pure dark matter and hydrodynamical constrained simulations of the Local Universe, aiming mostly at studying a variety of issues concerning the Local Group \citep[for a general review][and references therein]{2010arXiv1005.2687G}. Galaxy redshift surveys provide a different source of constraints. This was first pioneered by \citet{1998ApJ...492..439B} and followed later on by \citet{2002MNRAS.333..739M} and \citet{2010MNRAS.406.1007L} and very recently by \citet{2013MNRAS.435.2065H}. There is a considerable tradeoff between using peculiar velocities and spatial distributions of galaxies from redshift surveys. 

Galaxy redshifts are quite easy to measure accurately. Very large and deep surveys are now routinely produced. However, galaxy distributions constitute biased tracers of the underlying density field. The mass-to-light bias has yet to be completely modeled and corrected for.  
On the other hand, measuring peculiar velocities poses formidable challenges to observational cosmologists. The observations are susceptible to systematic biases, and the resulting catalogs are noisy, sparse and with an incomplete sky coverage. Still, on the theoretical side, peculiar velocities are unbiased tracers of the underlying mass distribution. As long as virial motions inside clusters can be suppressed, the construction of the underlying density and velocity fields can be easily performed.

The procedure of constraining ICs with peculiar velocities is based on the linear Wiener-Filter (WF) and Constrained Realization of Gaussian fields (CR) algorithms \citep[][]{1991ApJ...380L...5H,1995ApJ...449..446Z,2009LNP...665..565H}. The main deficiency of the method in these preliminary studies has been not to account for the cosmic displacement field of the data points from their primordial positions. Consequently, constrained simulated halos were, at z=0, located 10 \hMpc\ away from the reference objects' original positions. An effective remedy to the problem is to evaluate the Zel'dovich linear displacement field from the WF reconstructed overdensity field. Then, this displacement is reversed in time to move the constraints from z=0 backwards to their progenitors' positions at higher redshifts. This Reverse Zel'dovich Approximation (RZA) has been previously devised and tested against simple mock catalogs. For these simple mocks it was shown to recover positions up to a few Megaparsecs, typically $\approx 6 \hmpc$ \citep{2013MNRAS.430..902D,2013MNRAS.430..912D,2013MNRAS.430..888D}. 

The present paper takes three additional steps: 1) the technique to produce constrained ICs from peculiar velocities is refined, 2) the technique is tested on more realistic mocks which include a zone of galactic extinction and errors on distances, 3) the technique is applied to cosmicflows-1, a catalog of observational peculiar velocities \citep{2008ApJ...676..184T}.\\

The structure of the paper is as follows. Methodology and its refined version are described in the first section. Then section 2 presents the construction of realistic mocks drawn from a previous constrained simulation of the Local Universe. Resulting mock universes are presented and analyzed in section 3. This is followed by the application of the method on {a catalog of observed peculiar velocities}. Constrained simulations are analyzed in the fourth section. A general discussion concludes the paper.

Throughout this paper, distances are in \hMpc . All figures are presented after a gaussian smoothing of the fields at 2 \hMpc, which is the RZA technique intrinsic floor value of validity, as it will be re-measured in this paper.

%%%%%%%%%%%%%%%%%%%%%%%%%%%%%%%%%%%%%%%%%%%
% The method%%%%%%%%%%%%%%%%%%%%%%%%%%%%%%%%%%%%%%%%%%%
%%%%%%%%%%%%%%%%%%%%%%%%%%%%%%%%%%%%%%%%%%%

\section{Methodology: Constrained Initial Conditions}

The Constrained Realization technique is a very efficient and straightforward method to set ICs. It requires only the computation of a correlation matrix and its inverse. Assuming that the observed peculiar velocities are not strongly affected by non-linear dynamics (curl free field above the scale of virial motions), and assuming a prior cosmological model (here growth rate constant with time), the ICs are readily calculated. Detailed equations are available in Appendix A. 

The major drawback of the CR method is the fact that it is formulated in an Eulerian way: the cosmic displacement field is neglected, although galaxies observed today are at different comoving positions from their progenitors at higher redshifts. A first attempt to improve this shortcoming has been recently suggested by \citet{2013MNRAS.430..902D,2013MNRAS.430..912D,2013MNRAS.430..888D}: constraints are re-located at their precursors' positions before feeding them to the CR method. The progenitor's positions are computed with the quasi-linear Zel'dovich approximation using the cosmic displacement field provided by the WF estimator \citep[see Appendix B, for the equation and][for a detailed description]{2013MNRAS.430..902D}. However, because observed constraints have uncertainties, the peculiar velocity field is not accurately described by solely one (radial) component. An additional step can be added to the initial technique called RZA-radial from now on. In the refined technique (RZA3D) constraints are not only moved to their progenitors' positions but also the \emph{observed uncertain} peculiar velocities are replaced by fully WF-reconstructed three component vectors. Such resulting constraints have been given by the WF a weight according to their precision. The WF field goes to the null value when there is no coherent signal or when data points have too large errors (see Appendix A). Thus no error should be given to RZA3D derived constraints when input in the CR. ICs are then produced in the standard way, namely a random component and  the peculiar velocity constraints are combined to produce a primordial perturbation field assuming a prior power spectrum.

\begin{figure}
\hspace{-0.2cm}\includegraphics[width=0.5\textwidth]{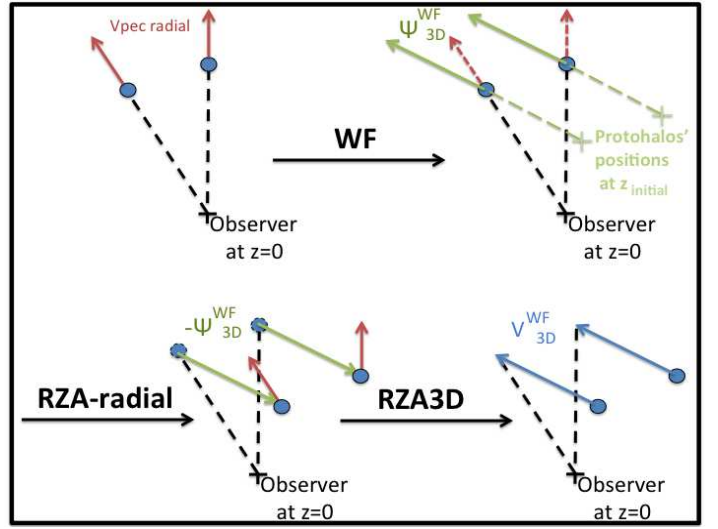}
\caption{The refinement of the  WF/RZA technique. The Wiener-Filter applied to observed radial peculiar velocities provides full 3D reconstructed peculiar velocity field $v^{WF}_{3D}$ which allows to derive the cosmic displacement field $\Psi^{WF}_{3D}$ (see Appendix B). In the initial RZA-radial technique the observational (radial) constraints at z=0 are re-located by -$\Psi^{WF}_{3D}$ to their progenitors' positions at higher redshifts. Since the peculiar velocity field is curl free (above the scale of virial motions)  it is supposed to be fully defined by only one component. However, because observed peculiar velocities have uncertainties, RZA-radial is insufficient. The proposed refinement in RZA3D takes care of this flaw by using the full 3D WF reconstructed peculiar velocities $v^{WF}_{3D}$ as constraints. }
\label{RZA3D}
\end{figure}

 In other words RZA3D differs from the initial RZA-radial on two points: 1) instead of observed radial peculiar velocities, the constraints are now the WF estimated peculiar velocities, 2) ICs are constructed under the assumption of null statistical errors (to prevent the double signal suppression resulting from the successive application of the WF (to obtain 3D velocities) and CR (to produce ICs)). Figure \ref{RZA3D} provides a schematic presentation of the WF/RZA method which prepares the constraints to be input in the CR algorithm. 
 
In the next section, RZA3D is applied to mock catalogs and is compared with the previous RZA-radial method. In both cases, a $\sigma_{NL}$ term accounts for the non-linear contributions of the radial peculiar velocities, which are not included in the model. It is added to the theoretical data-data correlated matrix \citep[cf. Appendix A and][]{2013MNRAS.430..902D}.

%%%%%%%%%%%%%%%%%%%%%%%%%%%%%%%%%%%%%%%%%%%%%%%%%%%%%%%%%%%%%
%%%%%%%%%%%%%%%%%%%%%%%%%%%%%%%%%%%%%%%%%%%%%%%%%%%%%%%%%%%%%
%Mocks: Testing the method%%%%%%%%%%%%%%%%%%%%%%%%%%%%%%%%%%%%%%%%
%%%%%%%%%%%%%%%%%%%%%%%%%%%%%%%%%%%%%%%%%%%%%%%%%%%%%%%%%%%%%
%%%%%%%%%%%%%%%%%%%%%%%%%%%%%%%%%%%%%%%%%%%%%%%%%%%%%%%%%%%%%

\section{Building Mock Catalogs}
\label{sec3}

When working on scales of a few tens of Megaparsecs (enclosing a volume we call the Local Universe), the cosmic variance is a major concern because these scales are far smaller than the scale of homogeneity of the Universe. As a matter of fact, the observational dataset used in this paper only reaches once to twice the size of our filament (from Ursa-Major to Centaurus clusters, the distance is roughly $40 \hmpc$). This leads to two biases. Firstly, our position (as observers) is unique (peculiar) and impacts highly the data collections. In other words since we are living in a supercluster bounding a very large void, the observed peculiar velocities are very much dominated by a specific local structure dynamics. 
Secondly, we are only observing one possible (coherent with the constraints) realization of the universe. Thus we must 1) test the methodology against a realistic replica of the actual nearby Universe to be in the same particular dynamical conditions (including the presence of a nearby very large void) and 2) run several constrained numerical simulations to estimate the confidence level at which the observed large scale structures are recovered with a given methodology whatever the random component.

Accordingly, mock catalogs are drawn from a previous constrained simulation of the Local Universe \citep{2003ApJ...596...19K}. This particular simulation, hereafter BOX160, is a dark matter only simulation of $1024^3$ particles in a computational box of side length $L=160 \hmpc$ constrained by peculiar velocity catalogs and a sample of positions  and masses of X-ray selected nearby clusters. Since this simulation was computed in the WMAP3 framework, tests on the mock are also conducted in the WMAP3 framework. WMAP3 is a flat universe with a matter density of $\Omega_m=0.24$, a $\sigma_8=0.75$ normalization and $H_{0}=73$ km s$^{-1}$Mpc$^{-1}$.

Box160 reproduces many of the key structures of the nearby universe, such as Virgo, Coma and Centaurus clusters, Perseus-Pisces supercluster and the Great Attractor region. A Local Group-like structure has been identified in the simulation and a mock observer is attached to that object. The catalog is built with respect to this observer. We assume that galaxies follow the peculiar velocities of dark matter halos in which they reside. A mock catalog of dark matter halos has been extracted from BOX160 with the Amiga halo finder \citep{2009ApJS..182..608K}. The output list of parameters contains halo coordinates in \hMpc, peculiar velocities in \kms\ and masses in h$^{-1}$ M$_{\odot}$. Halos are selected in a sphere of 30 \hMpc\ radius around the mock observer to mimic as much as possible the extent of the Cosmicflows project's first catalog to be used later on in this paper. Two novelties with respect to the mocks of  \citet{2013MNRAS.430..888D,2013MNRAS.430..902D,2013MNRAS.430..912D} are introduced. To simulate a zone without data similar to that produced by the extinction of our galaxy's disk (Zone of Avoidance), every halo with a latitude in between $\pm10^\circ$ is removed. Major players in the local dynamics, such as the mock Great Attractor, are thus (partly) masked by this extinction zone. The mock catalog is also designed to reproduce the current observational limits: a 20\% uncertainty on galaxy distances, and thereby on derived radial peculiar velocities. For simplicity the relative errors are assumed to follow a normal distribution around the true distances $d$. Resulting distances $d'$ correspond to the observational measured distances with their uncertainties. The mock radial peculiar velocities are then computed as the difference between the redshifts and the modified distances $d'$ multiplied by 100 (Hubble Constant $\times$ h$^{-1}$ as distances are in \hMpc). This procedure takes place in the Cosmic Microwave Background frame of reference, namely in the framework of the computational box. 

The reconstruction of the Large Scale Structure, and of its ICs, from peculiar velocities is hampered by virial motions of galaxies in clusters. Such motions cannot be accounted for in the present proposed WF/RZA/CR framework. Grouping distance measurements of galaxies at the same distance (belonging to a same cluster) into a single data point with a reduced error could provide a partial remedy to the problem. However, the issue of grouping a given catalog constitutes a formidable challenge that has not been adequately solved yet. Fortunately late-type galaxies constitute roughly 80\% of cosmicflows-1 data \citep{2008ApJ...676..184T}. Such galaxies reside in the field where virial motions do not dominate cosmic flows, thus they are less affected by this problem. The mock catalog is quite similar on this matter. The halo selection procedure described earlier give 1467 halos. Still, 95\% of these halos that serve as mock data points are isolated. As a result, virial motions do not dominate cosmic flows in the mocks either. Such mock catalog is a reasonable proxy to the first catalog of the project Cosmicflows used in the second part of this paper.

%%%%%%%%%%%%%%%%%%%%%%%%%%%%%%%%%%%%%%%%%%%
%Applications%%%%%%%%%%%%%%%%%%%%%%%%%%%%%%%%%%%%%%%%%%%
%%%%%%%%%%%%%%%%%%%%%%%%%%%%%%%%%%%%%%%%%%%

\section{Constrained Simulations with mock data}

The full machinery to obtain constrained simulations is tested on the mock built in section \ref{sec3}. The WF/RZA/CR algorithm is first applied to obtain ICs. These latter are then input in \gadgettwo\  \citep{2005MNRAS.364.1105S} N-body code to perform dark matter only simulations. The outcomes are compared with the initial BOX160. The boxsize, 160 \hMpc\ long on each side, is almost three times the extent of the mock. Periodic boundary conditions can be assumed without any risk of spurious phenomena in the central 60 \hMpc\ region to be analyzed. The grid size is $N=256^3$. To avoid shell-crossing, simulations are started at z=60. They are run until z=0. 

%%%%%%%%%%%%%%%%%%%%%%%%%%%%%%%%%%%%%%%%%%%
\subsection{Wiener-Filter reconstruction of the mock universe}

\begin{figure}
\centering
\includegraphics[width=0.45\textwidth]{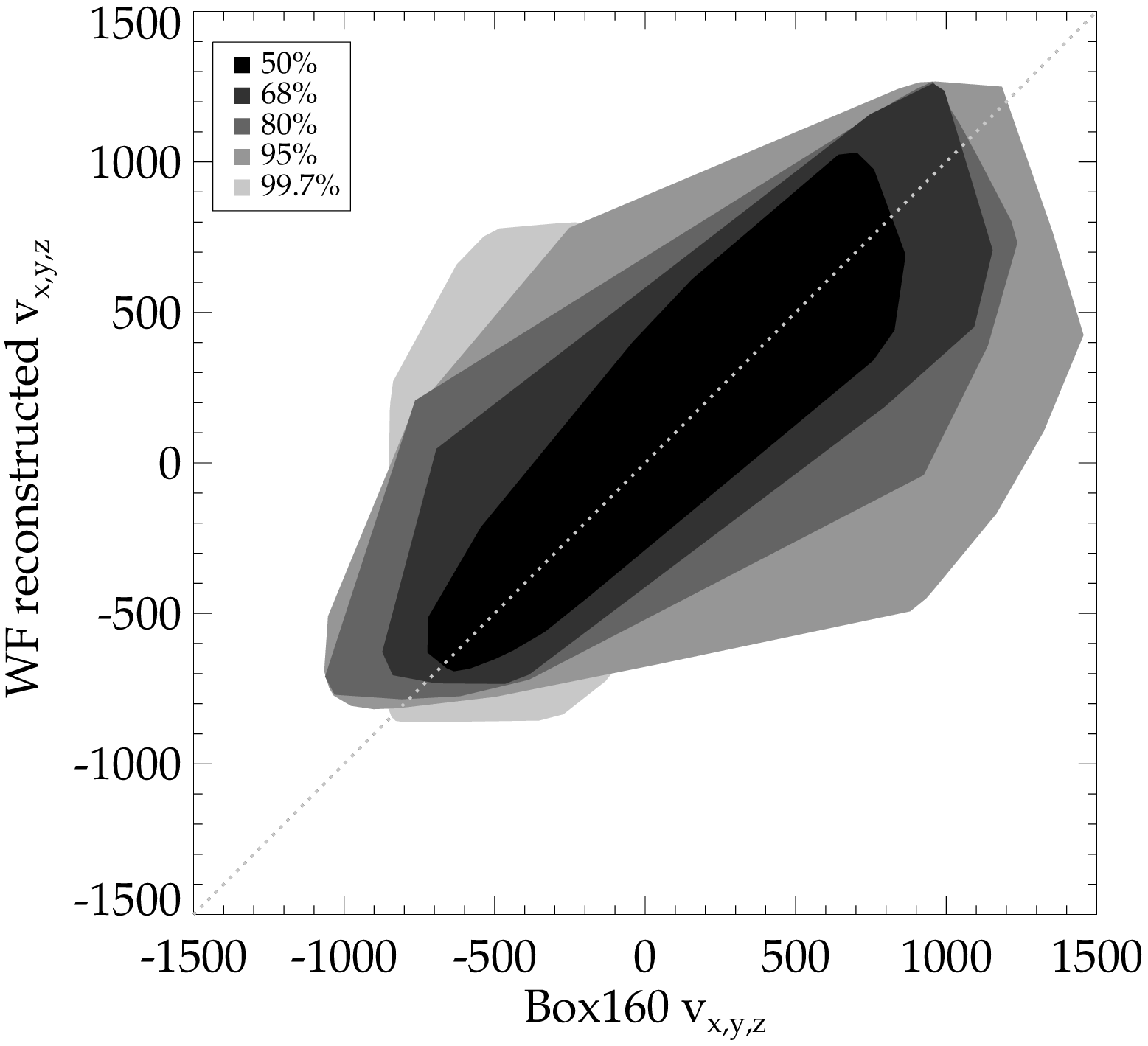}
\caption{Confidence level zones obtained with a cell-to-cell comparison between the velocity grids of the reference simulation BOX160 and its WF reconstruction within a 30 \hMpc\ radius sphere. For example, 50 \% of the peculiar velocity pairs, (value$_{cell \;xyz \; BOX160}$ ; value$_{cell\;xyz\;WF}$) can be found in the darker zone. This zone, slightly scattered around a 1:1 relation, shows the good agreement between the velocity values in a cell from the reference simulation and in the very same cell from the reconstruction.  68 \%  of the pairs are in the sum of the two internal zones and so on. 68, 95 and 99.7 \%  correspond to 1, 2 and 3-$\sigma$ uncertainties.}
\label{compc2c}
\end{figure}

To facilitate the comparisons between the initial BOX160 and its WF reconstruction from a mock, the BOX160 velocity field is interpolated on a $256^3$ grid using a Clouds-In-Cells interpolation scheme. Figure \ref{compc2c} displays the confidence level zones of the reconstruction with respect to the original BOX160. The zones result from a cell-to-cell comparison, within the central 60 \hMpc\ region, between the velocity grids of the WF reconstruction and of the reference simulation. The total scatter around the 1:1 relation is 201\kms\ or 1.7 $\sigma$. The WF velocity field is thus a good reconstruction of the BOX160 source.

\begin{figure*}
\centering
\vspace{-0.5cm}
\hspace{0.2cm}\includegraphics[width=0.76\textwidth]{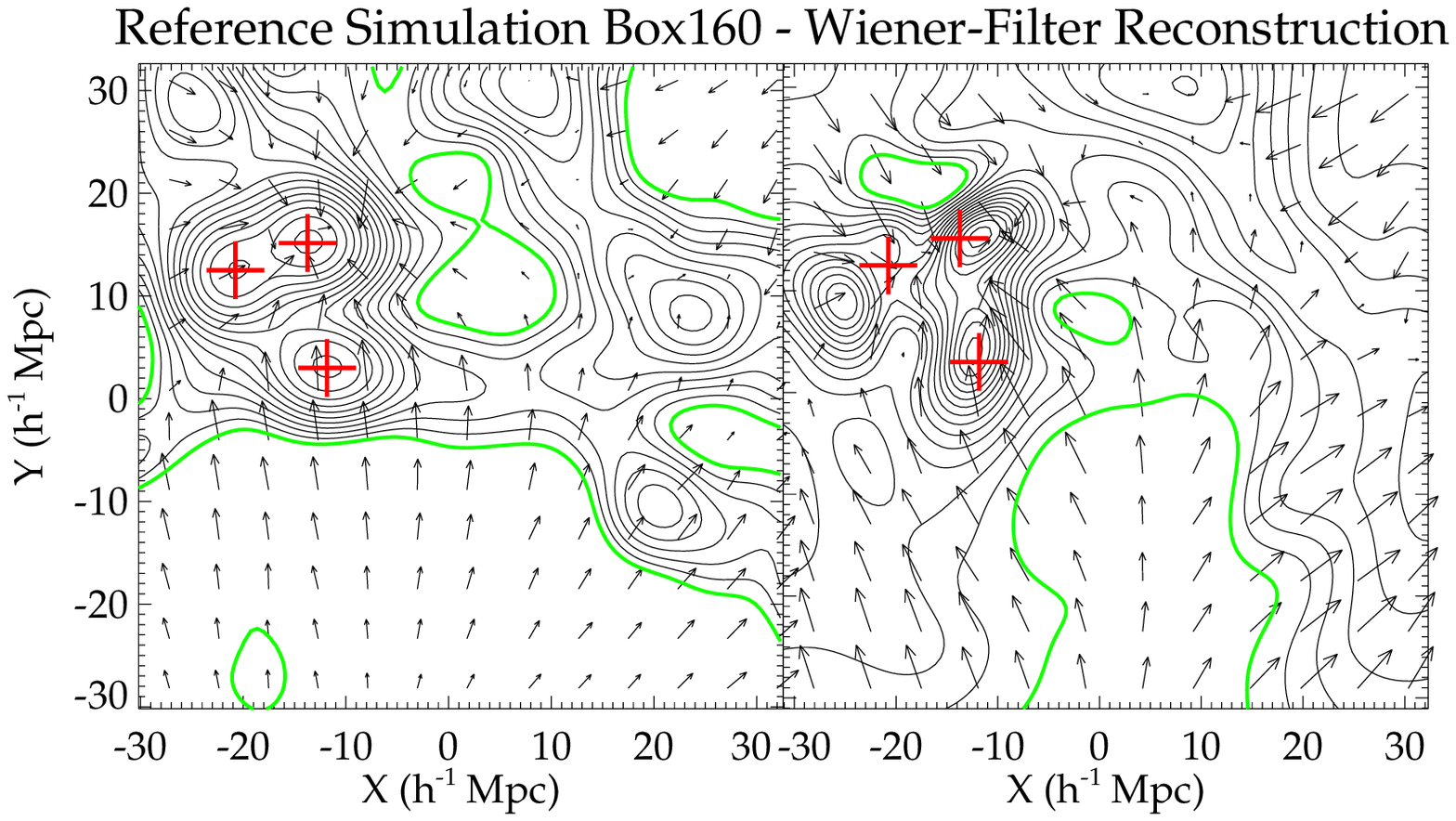}\\
\vspace{0.5cm}
\includegraphics[width=0.75\textwidth]{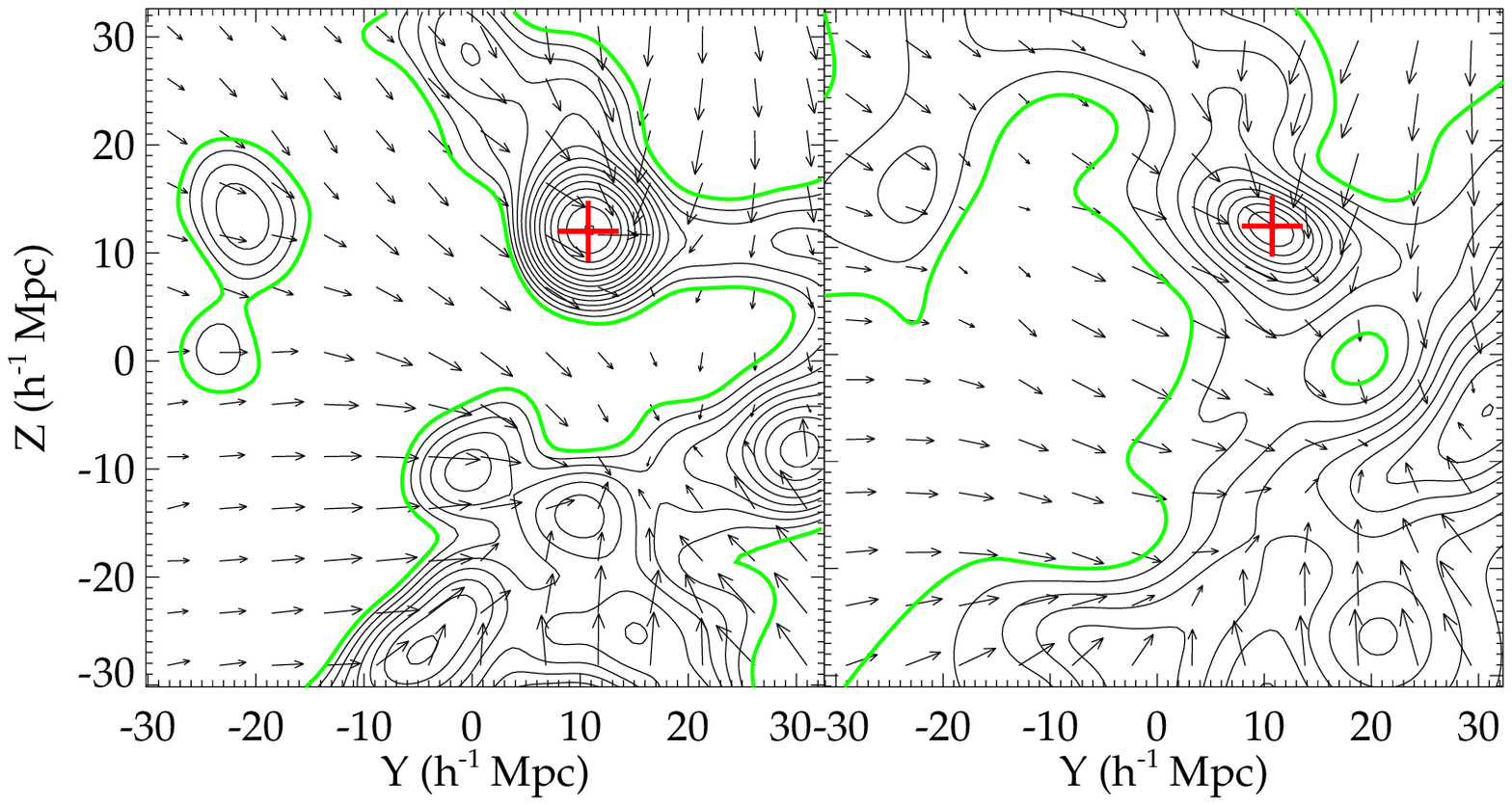}\\
%\vspace{0.5cm}
%\includegraphics[width=0.75\textwidth]{mock_256_160box160_256_Timurbox160_128_TimurSGXZ00.eps}%box160_256_TimurZ_SGXYZ.eps}
\caption{XY and YZ views of the reference simulation BOX160 (left) and its Wiener-Filter reconstruction (right) restricted to the central 60 \hMpc\ zone. The reconstruction has been obtained using only radial peculiar velocity data from a realistic mock catalog containing about the same errors and same number of data-points as the observational cosmicflows-1 catalog. The cosmic flows are represented by black arrows. Overdensity iso-contours are delimited by solid black lines. The green contour delimitates the mean density. Red crosses show the positions of the major density peaks in the reference simulation. Even with this sparse and noisy realistic mock, the WF has enough signal to properly recover the original cosmic flows and density peaks, with a precision of about 2-3 \hMpc\ in position.}
\label{box160WF}
\end{figure*}

Figure \ref{box160WF} shows two planes centered on the look-alike of the Milky-Way of the reference simulation and of its reconstruction obtained with the WF applied to the mock. Velocity (black arrows) and density fields (contours) are plotted. The green contour displays the mean density level. The main features - direction of the cosmic flows and attractors' positions - are properly reconstructed. The feature in the XY plane is the Great Attractor region look-alike with three density peaks from the reference simulation marked by red crosses in both quadrants. In YZ the red cross locates the density peak of the mock Virgo halo in the reference simulation. These qualitative analyses illustrate the claim that with a sparse and noisy mock similar to cosmicflows-1 (in terms of number of constraints, zone of extinction without data and large errors on peculiar velocities) the WF is an optimal reconstruction tool in the linear regime of the gravitational instability. Structures are not necessarily reconstructed at their exact positions since the intrinsic accuracy is about 2 \hMpc . Still, overall, the density field is recovered when considering only the linear theory on all scales.

%%%%%%%%%%%%%%%%%%%%%%%%%%%%%%%%%%%%%%%%%%%
\subsection{Constrained Simulations: RZA-radial versus RZA3D}

Once the continuous fields obtained with the WF technique are extrapolated at the data points' positions, constraints are displaced from their z=0 location to their progenitors' position at higher redshifts. In addition, constraints are replaced by their full WF reconstruction in the RZA3D technique. Since simulations are run with periodic boundary conditions, only the divergent part of the velocity field (velocities due to densities inside the box solely) is used to generate ICs. Hence, any tidal motion due to densities outside of the box is removed.

\begin{figure}
\hspace{-1cm}\includegraphics[width=0.6\textwidth]{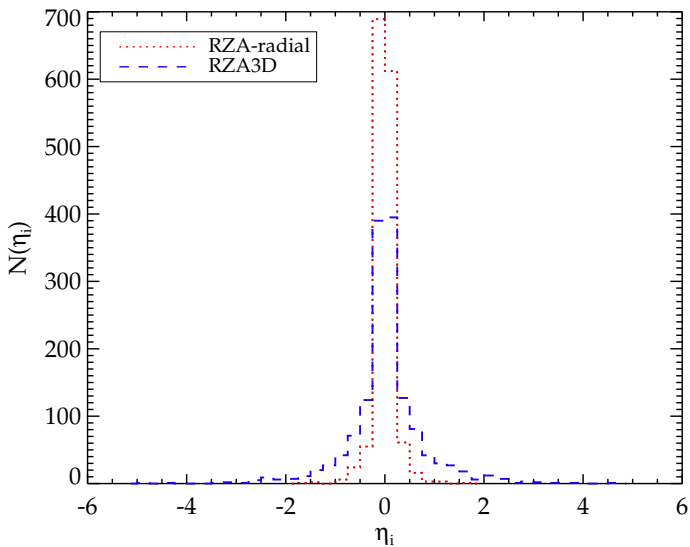}
\vspace{-2cm}
\caption{Distribution functions of the component values $\eta_i$ of the data-data correlation vector $\eta$ computed with RZA-radial (dotted red line) and RZA3D (dashed blue line) constraints. The higher the absolute value of $\eta_i$, the more the corresponding constraint contributes in the constrained Initial Condition. Since the distribution obtained with RZA3D is wider than that resulting from RZA-radial, RZA3D constraints are stronger than RZA-radial ones.}
\label{corrvec}
\end{figure}

A major objective of the paper is to compare RZA-radial and RZA3D algorithms. However, cosmic variance can affect the comparison between methods. To take care of this effect, 1) each RZA-radial derived Initial Condition shares the same random component with one of the RZA3D obtained Initial Condition. Hence, the simulations resulting from the same random seed ICs are expected to reproduce the same Large Scale Structure ; 2) ten constrained ICs are built to estimate the confidence level on structure position for each procedure. Resulting simulations are also compared with the reference BOX160 to estimate the average misplacement of simulated structures at z=0 with respect to original locations. The comparison between the constrained simulations and BOX160 is done on a $256^3$ Clouds-In-Cells grid after smoothing the density and velocity fields with a Gaussian kernel of $2.0\, \hmpc$. When averaging over an increasing number of constrained simulations, the standard deviation with respect to BOX160 starts at 0.47 in logarithmic unit of density for one simulation and decrease to a plateau value of 0.37 when considering 8 or more simulations. Adding more than 10 simulations would not produce on average other high and deep density zones that could be compared between the two methods and with BOX160 (or the 0.37 value would have continued to decrease).
%Table \ref{table00} shows that 
The standard deviation of RZA3D simulations around their average is smaller than that of RZA-radial both in terms of velocity and density (0.34 against 0.35 in logarithmic unit of density and 246 against 258 \kms). Although there is a random component, constrained simulations of BOX160 obtained with RZA3D method have stronger features reproduced at very similar positions than RZA-radial constrained simulations. The cosmic variance is reduced with RZA3D because constraints are stronger than in the RZA-radial case as seen on Figure \ref{corrvec}. The $\eta_i$ components of the data-data correlation vector $\eta$ have higher absolute values with RZA3D than with RZA-radial (see Appendix A for detailed equations).

%The degree by which the ICs are constrained depends on the quality of the data, of the assumed prior model and on the dynamic range of the constraints. 

%\begin{table}
%\begin{center}
%\begin{tabular}{|c|c|c|c|c|}
%\hline
%Method on Mock & $\sigma_d$ & $\sigma_{vx}$ & $\sigma_{vy}$ & $\sigma_{vz}$\\
%\hline
%RZA-radial & 1.59 & 167 & 168 & 178 \\
%RZA3D & 1.44 & 162 & 160 & 167 \\
%\hline
%\end{tabular}
%\end{center}
%\caption{Standard deviations of the 10 simulations around their average using the mock catalog: (1) Method used to generate ICs. (2) Variance around the average density field. (3), (4) and (5) Variance around the x, y and z components of the average velocity field in \kms\ respectively.}
%\label{table00}
%\end{table}

%Figure \ref{massfunction} shows the dark matter halo mass functions of the 10 RZA3D constrained simulations. The mass function of 10 random simulations, drawn from the same power spectrum, are shown for reference (red area). The constrained mass functions exceed the random ones at high mass, namely the constrained simulations have more massive halos than $10^{14}\hmsun$ than the random ones.
%\begin{figure}
%\centering
%\includegraphics[width=0.48\textwidth]{mockcf1_RZA3D_256_160massfunction_WMAP3.eps}
%\caption{Mass Functions of the 10 RZA3D simulations constrained with the mock built from BOX160 (black dashed lines). The red shaded area is the range within which the 10 random simulations built with the same cosmological parameters can be found. At high masses, the mass functions of re-simulations are slightly above the range of the random.}
%\label{massfunction}
%\end{figure}

\begin{figure*}
\centering
\vspace{-1.7cm}
\includegraphics[width=0.48\textwidth]{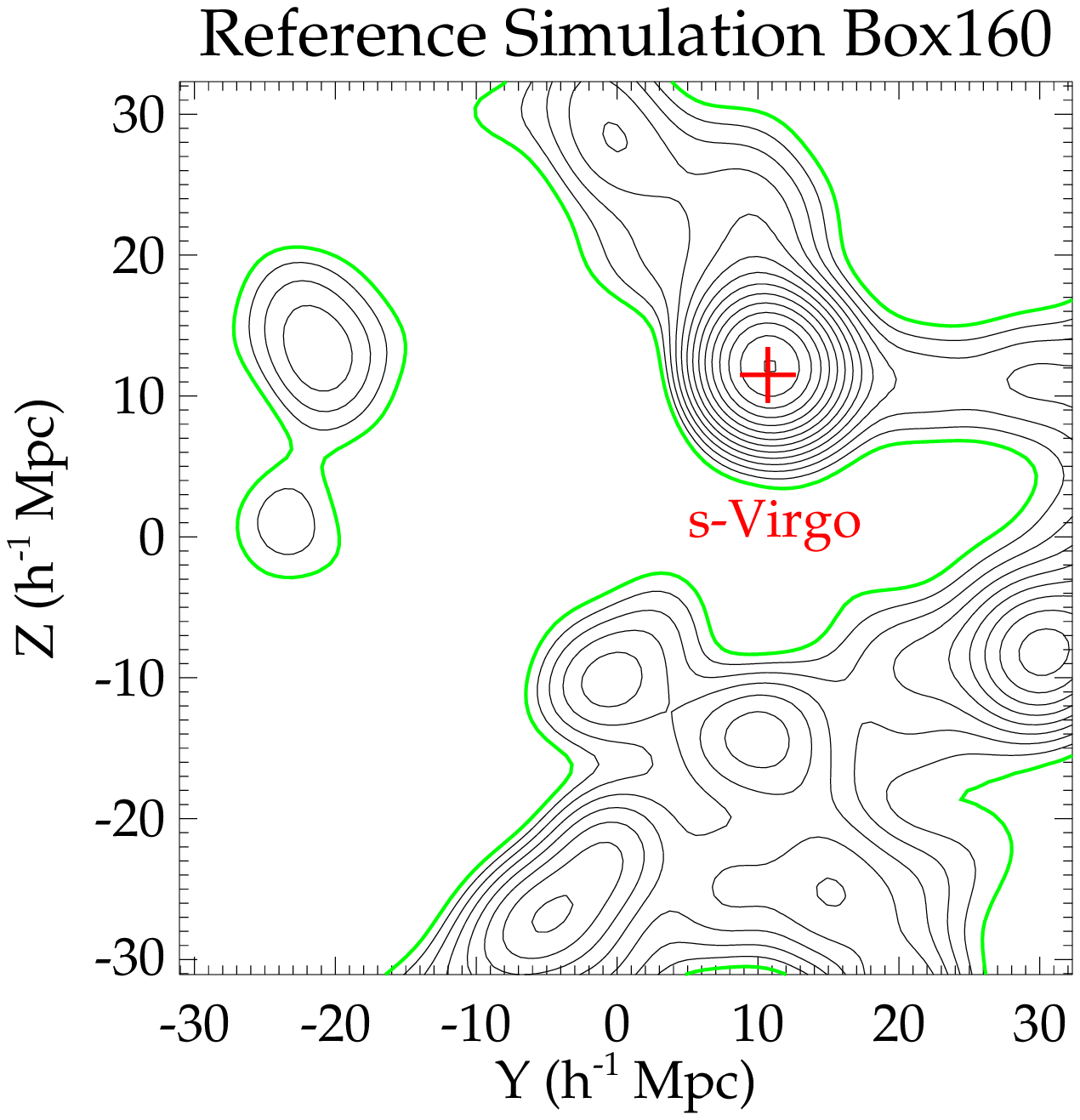}
\includegraphics[width=0.48\textwidth]{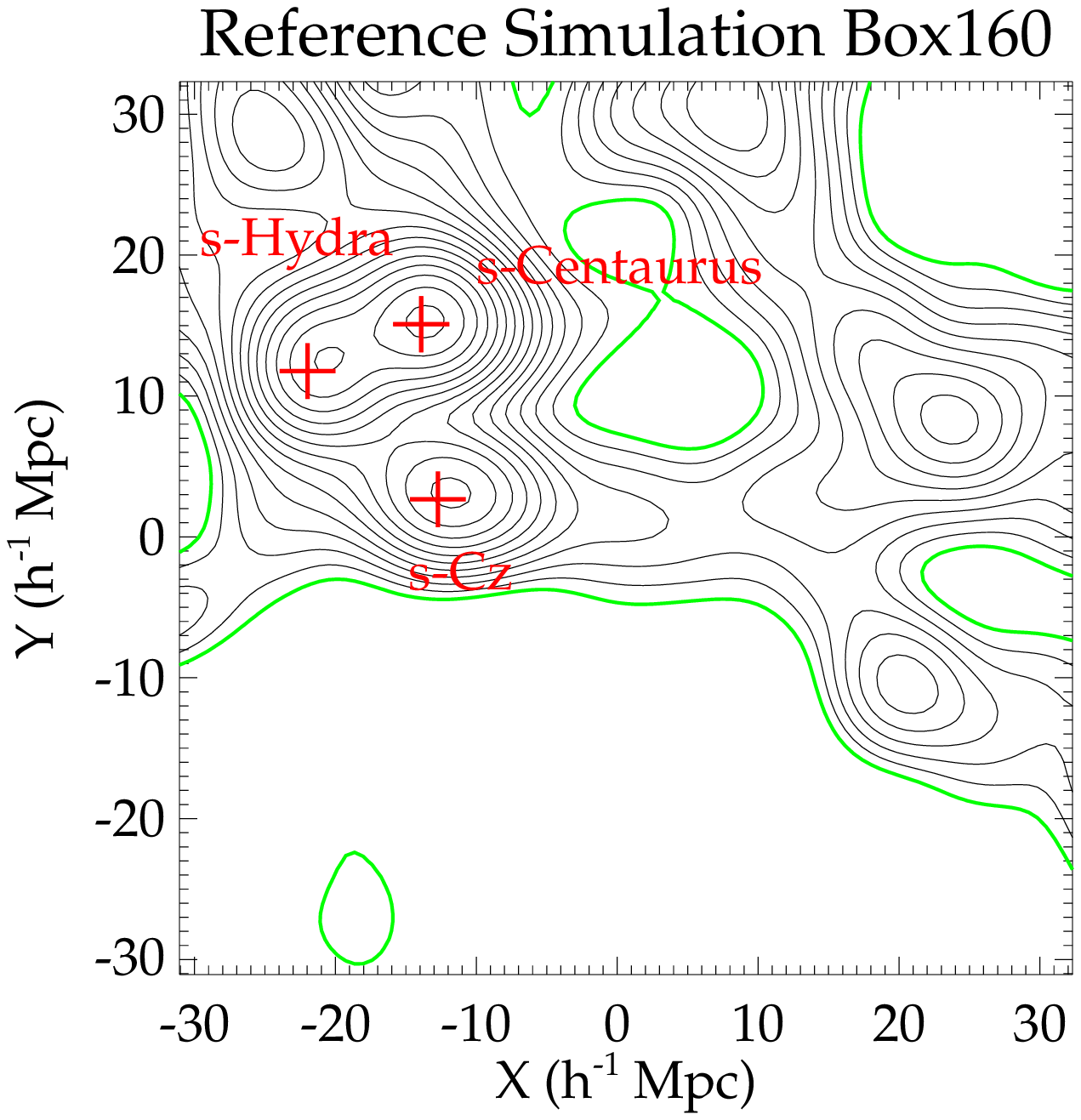}\\
\vspace{-1cm}
%box160_256_TimurshiftY__SGX=6pt83495_cont_vel.eps}\\
\includegraphics[width=0.48\textwidth]{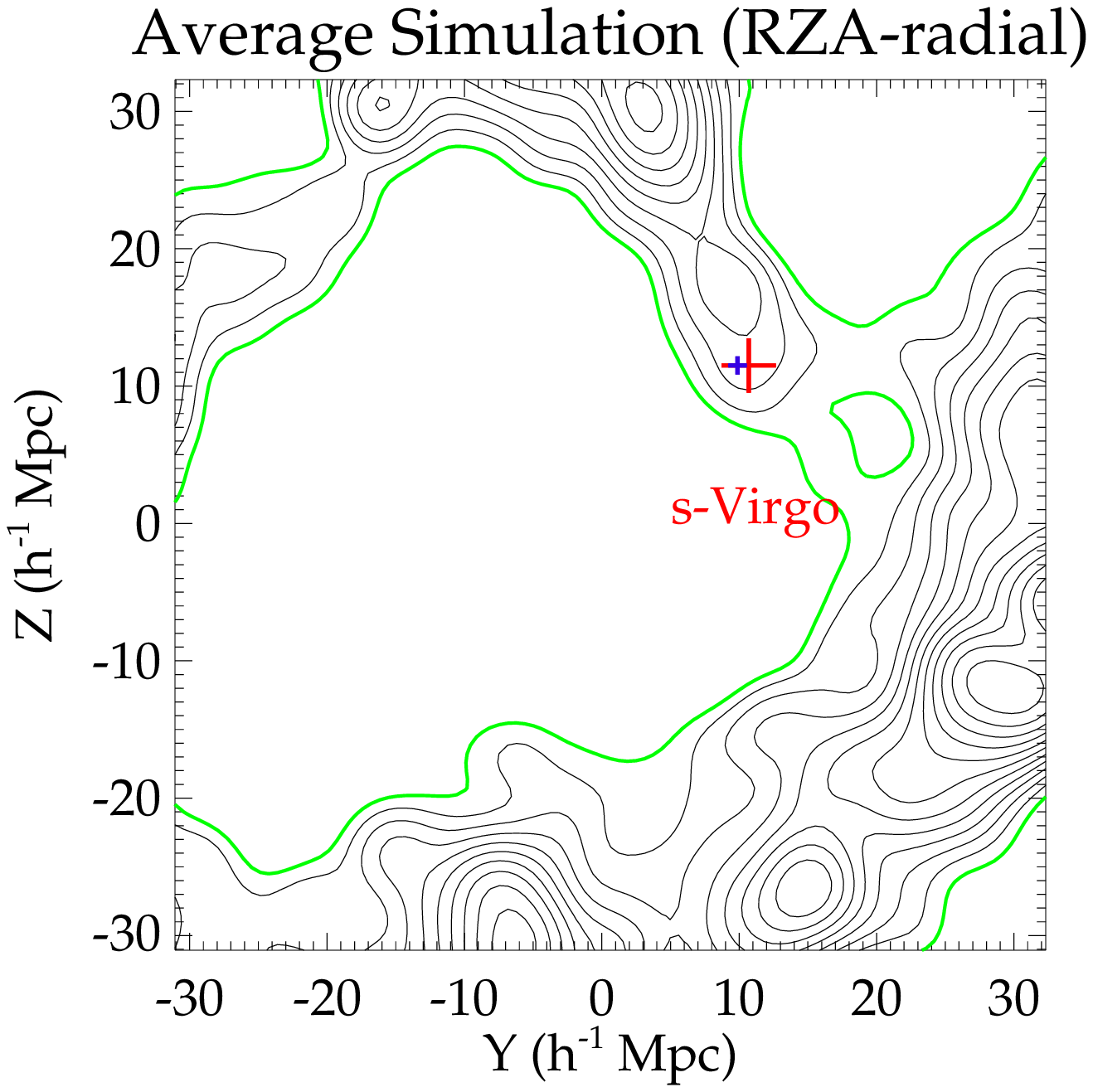}
\includegraphics[width=0.48\textwidth]{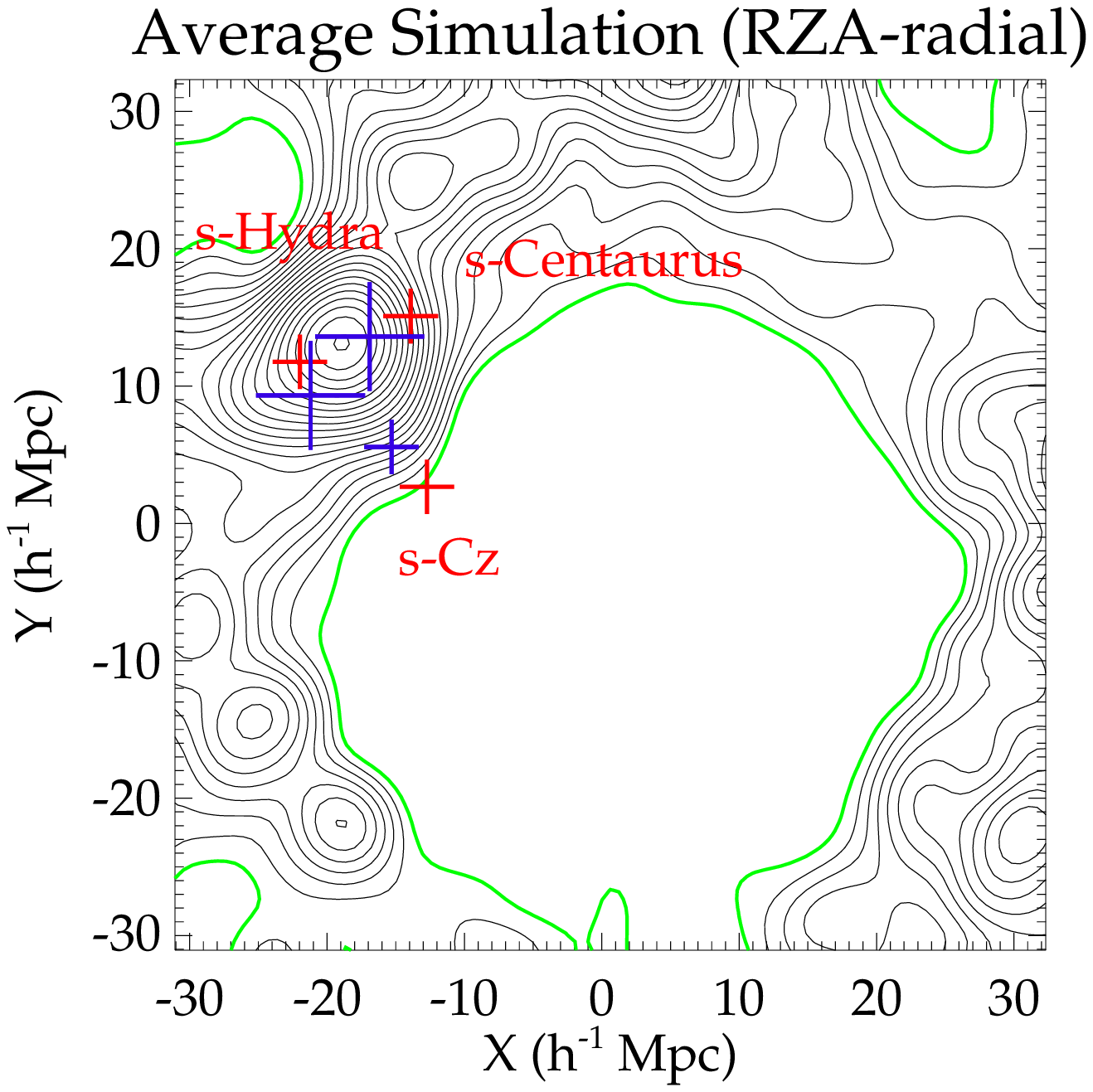}\\
\vspace{-1cm}%mockcf1_RZA_256_160shiftY__SGX=6pt83495_cont_vel.eps}\\
\includegraphics[width=0.48\textwidth]{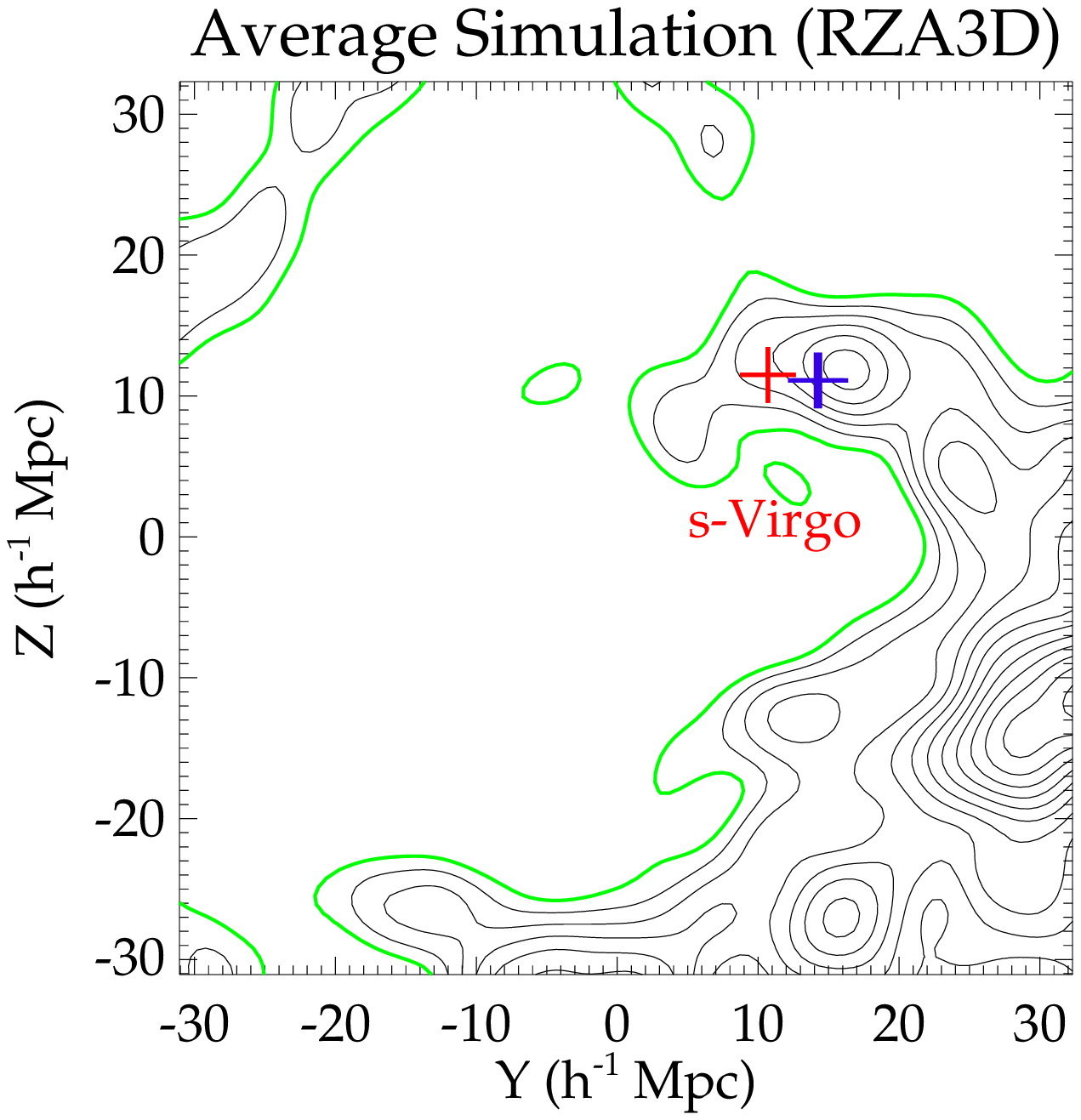}
\includegraphics[width=0.48\textwidth]{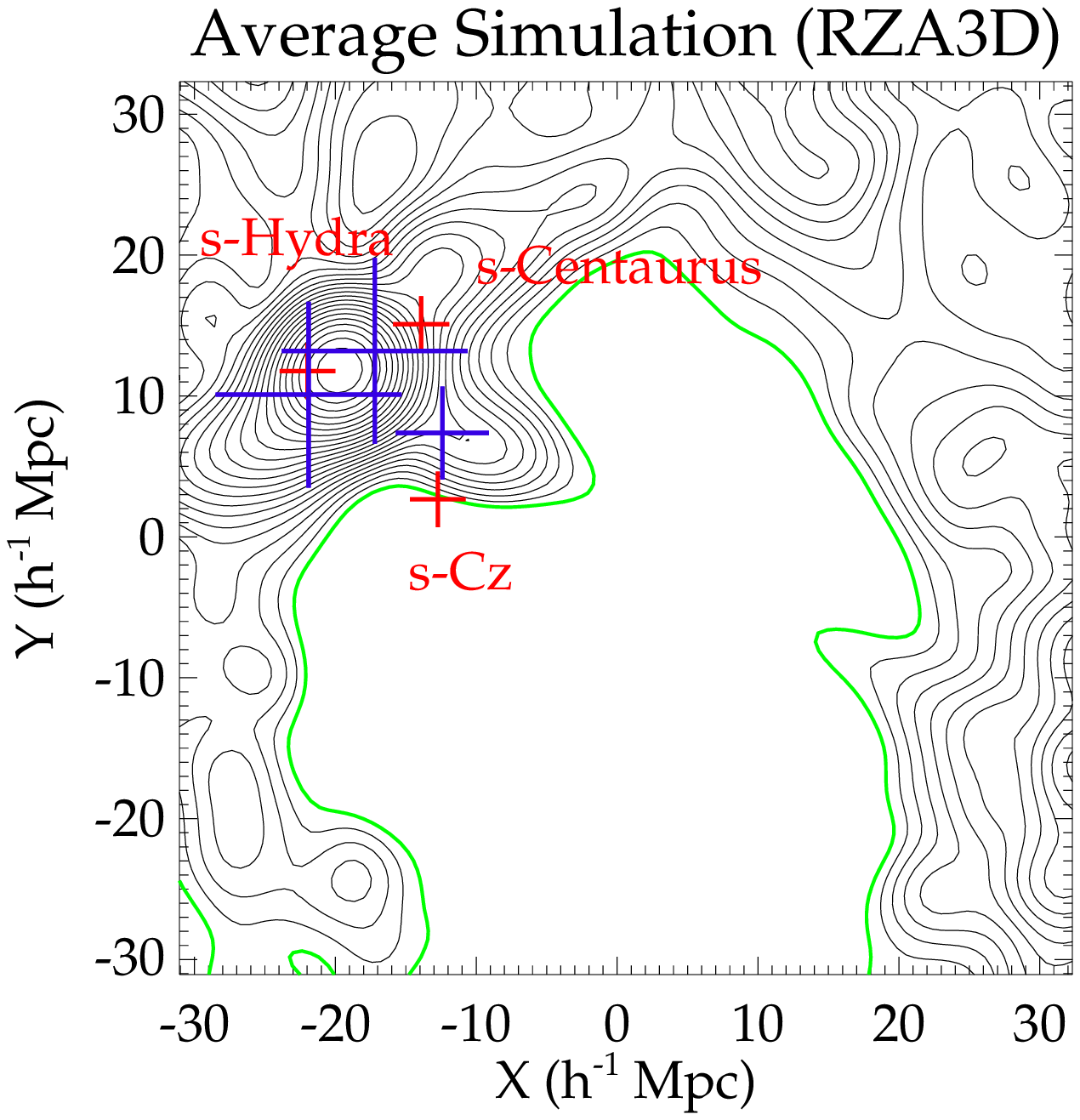}
\vspace{-0.5cm}
%mockcf1_RZA_256_160shiftY__SGZ=-6pt21359_cont_vel.eps}\\
\caption{Visualization of planes containing main simulated attractors (positioned at X=7 and  at Z=-6 \hMpc ). Solid black iso-contours delimit overdensities. The green color stands for the mean density in the box. Top: Reference simulation. Middle: Average over 10 constrained simulations using RZA-radial on the mock. Bottom: Average over 10 constrained simulations applying RZA3D on the mock. Red crosses show original positions of s-Virgo, s-Hydra and s-Centaurus in the reference simulation. Positions of the averaged replicas in the constrained simulations are shown in blue.  Crosses' sizes are proportional to the number of replicas found out of 10 simulations.}
\label{Virgo-Centaurus}
\end{figure*}

BOX160 contains some replicas of prominent nearby structures such as Virgo, Hydra and Centaurus. These halos are named hereafter s-Virgo, s-Hydra and s-Centaurus to distinguish them from the observed ones. BOX160 contains also a halo called s-Cz \citep[in accord with][]{2013MNRAS.430..912D}. These target objects are used to monitor the quality of the simulations. Figure \ref{Virgo-Centaurus} shows the density field in the planes containing these objects of the actual simulation BOX160 (top panel), of RZA-radial (middle panel), and RZA3D (bottom panel) with simulations averaged on 10 different realizations. The main dark matter halos from BOX160 used as tracers are marked by red crosses in the six panels. Only in the RZA3D simulations there is a recurrent overdensity at the expected location of s-Virgo (high density peak in the YZ plot).

For each one of the simulations, dark matter halos are obtained with the Amiga halo finder and the s-halos are identified when recovered. A halo in a constrained simulation is considered to be a replica of a BOX160 halo when the difference in position is smaller than $\sim$6 \hMpc\ and when masses are of the same order. The search is restricted to a sphere of $\sim$6 \hMpc\ since the scope of this work is to find a method resulting in an error below 6 \hMpc\ (3 $\sigma$). Blue crosses on Figure \ref{Virgo-Centaurus} are located at the average position of the look-alikes of s-Virgo, s-Hydra, s-Centaurus and s-Cz halos in the constrained simulations. The cross sizes are proportional to the number of simulations (out of 10) in which a replica has been found. Table \ref{table2} recapitulates the characteristics of the targeted halos and of their look-alikes: virial masses, positions and standard deviations. RZA-radial fails to recover s-Hydra and s-Centaurus as separate individual objects in 5 out of 10 simulations,  thus they are not reported in the table. In these 5 out of 10 simulations, they are collapsed into a single object. The table also records the average of each replicaÕs distance to the genuine halo. The typical difference is about 5 \hMpc\ for the RZA3D technique against 6 \hMpc\ for RZA-radial. However, because with RZA-radial more halos are not found in the 6 \hMpc\ sphere (they are outside of the sphere so farther away) than with RZA3D, the value for RZA-radial is more biased (lowered) by the restricted search than that of RZA3D. Still, studying the standard deviations of the position errors shows that it is possible to reach the floor value imposed by the linear regime, 2-3 \hMpc\ with some random seeds for RZA3D while it is always impossible with RZA-radial. The table proves an enhanced accuracy of the RZA3D method in terms of position errors (when compared with BOX160). The gain is also clear in term of reliability-robustness of the results since more replicas (out of ten different random seed simulations) are found at a similar location (smaller standard deviations in positions) with RZA3D than with RZA-radial.

We can also consider a comparison between high density peaks in the WF and in the constrained simulations.  The density peak reconstructed by the WF in the bottom right quadrant of Figure \ref{box160WF} is also present in 6 RZA3D simulations out of 10 when looking within a $\sim$ 6 \hMpc\ sphere centered on the WF peak. By contrast, there is a peak in only 3 RZA-radial simulations out of 10 within the same sphere. In both cases, the typical misplacement is 4-5 \hMpc\ with a standard deviation about 1 \hMpc. RZA3D applied to a mock cosmicflows-1 catalog outperforms RZA-radial applied to the same mock. The stronger the constraints, the more the cosmic variance that exists over ten constrained simulations because of a different random component is reduced. The number and accuracy of constraints in a cosmicflows-1-like catalog are adequate to simulate properly a look-alike of the Local Universe with a precision reaching the intrinsic limitation of the technique.

\begin{table*}
\begin{center}

\begin{tabular}{lllll}
\hline
Simulation case & Mass & Average Position X, Y, Z & Average Distance to  & Nb of \\
& && reference halo &  occurences\\

\hline
\hline
Box160 s-Virgo &3.3 & 7.12, 10.7, 11.5 & &\\
\underline{RZA-radial} s-Virgo & 0.25 & 3.40, 9.88, 17.2 & \underline{6.9 }& \underline{1/10}\\
\fbox{RZA3D} s-Virgo &  0.34 ; $\sigma=0.05$ & 7.78, 14.4, 11.1 ; $\sigma$= 2.4 & \fbox{ 5.4} ; \fbox{$\sigma$=1.6} & \fbox{5/10}\\ %3.8 ; $\sigma$=2.4& 5\\
\hline
\hline
Box160 s-Centaurus & 6.07 & -13.9, 15.1, -8.81 & & \\
\underline{RZA-radial}  s-Centaurus& 3.1  ; $\sigma=2.8$ & -16.9, 13.6, -9.62 ; \underline{$\sigma$= 3.0}& \underline{6.0} ; \underline{$\sigma=0.4$} & \underline{5/10}\\%3.5 ; $\sigma=3$& 5\\
\fbox{RZA3D}  s-Centaurus & 7.9  ; $\sigma=4.0$ & -17.2, 13.2, -10.5 ; \fbox{ $\sigma$= 1.9}&  \fbox{ 5.0} ; \fbox{$\sigma=1.5$} & \fbox{ 10/10}\\ %4.0 ; $\sigma=1.9$&10 \\
\hline
 \hline
 Box160 s-Hydra & 5.18 & -22.0, 11.8, -3.35 &  & \\
 \underline{RZA-radial}  s-Hydra & 4.7 ; $\sigma=2.3$ & -21.2, 9.32, -4.87 ; \underline{$\sigma$= 1.9} & 3.1 ; $\sigma=1.5$ & \underline{5/10} \\%3.0 ; $\sigma=1.5$ & 5 \\
 \fbox{RZA3D} s-Hydra  & 4.7; $\sigma=2.0$ & -21.9, 10.1, -4.58 ; \fbox{ $\sigma$= 1.3} & \fbox{  3.0} ; \fbox{ $ \sigma=1.2$} & \fbox{ 10/10} \\
\hline
\hline
Box160 s-Cz & 0.96 & -12.7, 2.68, -6.36 & & \\
\hline
\underline{RZA-radial}  & 0.80 ; $\sigma=0.44$ & -15.3, 5.56, -10.1 ; $\sigma$= 2.3 & 6.3 ;  \underline{$\sigma=0.57$} & \underline{3/10}\\% 5.4 ; $\sigma=2.0$& 3\\
\fbox{RZA3D} & 0.40 ; $\sigma=0.19$ & -12.4, 7.37,  -7.20 ; $\sigma$= 2.3 & 6.0 ; \fbox{$\sigma=1.9$} &\fbox{ 5/10}\\ %4.8 ; $\sigma=2.3$&5\\
\hline
\end{tabular}
\end{center}
\caption{Average parameters and standard deviations $\sigma$ for target halos looked for in a 6 \hMpc\ sphere centered on their original positions in the reference simulation.  (1) simulation in which the halos are looked for, (2) dark matter mass in  h$^{-1}$ $10^{-14} $  solar mass, (3) average coordinates X, Y and Z  in \hMpc\ and standard deviations, (4) average distance in \hMpc\ to the genuine halo and standard deviation $\sigma$, (5) number of simulations (out of ten with a different random seed) which contain a replica.}
\label{table2}
\end{table*}

%%%%%%%%%%%%%%%%%%%%%%%%%%%%%%%%%%%%%%%%%%%%%%%%%%%%%%%%%%%%%%%%%%%%%%%%%%%%%%%%%%%%%%%%%%%%%%%%%%%%%%%%%%%%%%%%%%%%%%%%%%%%%%%%%%%%%%%%%%%%%%%%%%%%%%%%%
%CF1: observational dataset%%%%%%%%%%%%%%%%%%%%%%%%%%%%%%%%%%%%%%%%%%%%%%%%%%%%%%%%%%%%%%%%%%%%%%%%%%%%%%%%%%%%%%%%
%%%%%%%%%%%%%%%%%%%%%%%%%%%%%%%%%%%%%%%%%%%%%%%%%%%

\section{Constrained Simulations with Cosmicflows}

%%%%%%%%%%%%%%%%%%%%%%%%%%%%%%%%%%%%%%%%%%%
\subsection{WF reconstruction of the Local Universe}

The Cosmicflows-1 catalog of peculiar velocities \citep{2008ApJ...676..184T}, compiled within the framework of the Cosmicflows project \citep[e.g.][]{2011MNRAS.415.1935C,2011MNRAS.414.2005C,2012ApJ...749...78T,2012AJ....144..133S,2013ApJ...765...94S,2012ApJ...758L..12S}, is from now on used in this paper to perform constrained simulations of the Local Universe. For every simulation, we assume a $\Lambda CDM$ model in the  7-year Wilkinson and Microwave Anisotropy Probe (WMAP7) framework \citep{2011ApJS..192...18K}. The cosmological parameters are $H_0=70$ \kms\ Mpc$^{-1}$, $\Omega_m$=0.272 and $\Omega_{\Lambda}$=0.728. On Figure \ref{wfsimucf1} the left column is the first step of the WF/RZA/CR technique, namely the Wiener-Filter. The middle and right columns represent the outcomes of one single simulation and of the average over 10 different realizations as described in subsection \ref{subset:CS-CF1-WMAP7}. In the reconstruction, supergalactic cartesian coordinates are centered on the Milky-Way and the XY supergalactic plane contains the Local Structure. In the simulations, the supergalactic coordinates are parallel to the box coordinates with the origin at the center of the box where ideally there is a Milky-Way-like. The two main planes are shown at the supergalactic Z=0 and X=$-2.5$ \hMpc\ coordinates to fit the location of the Virgo cluster. For a direct comparisons between the real Universe and the WF reconstructed large scale structure the V8k galaxy redshifts catalog is overplotted as grey dots. This catalog contains 30,124 galaxies with distances modified by a numerical action model of the Virgo infall for $V<3000$ \kms. It is available at the Extragalactic Distance Database website \citep[http://edd.ifa.hawaii.edu,][]{2009AJ....138..323T}. Qualitatively, the real Virgo cluster and the void behind are reconstructed in the supergalactic XY plane. Virgo is also visible next to the Local Void in the YZ supergalactic plane. 

%Note however that there is a small shift exclusively in the SGX coordinate about 4-5 \hMpc\ between the reconstruction and the V8k redshift catalog probably because of the bulk flow in the direction of the Great Attractor and Shapley, outside of the box which creates a tidal contribution to velocities in that direction that are not modeled.\\

%%%%%%%%%%%%%%%%%%%%%%%%%%%%%%%%%%%%%%%%%%%
\subsection{Constrained Simulations of the Local Universe}

The WF reconstructed velocity field is extrapolated at the location of the data points. Then, RZA-radial and RZA3D methods are applied to cosmicflows-1 constraints in order to construct ICs. Ten random seeds are used to build ICs with periodic boundary conditions for both methods. It is to be noted that observational radial peculiar velocities are measurements of motions related to the whole gravitational potential. However, simulations are here run in a relatively small box with periodic boundary conditions where not all the attractors responsible for the entire motions are present. Thus, replacing velocities that result from the entire gravitational potential by velocities related to the gravitational potential in the box (namely RZA3D instead of RZA-radial) is more accurate and constitutes another advantage of the RZA3D method. %Properly, the constrained simulations should be based on a larger catalog of constraints to include a volume that encloses all the significant gravitational influences. 
Resulting ICs are then run from redshifts 60 to 0. The cosmic variance in terms of standard deviations of ten simulations with respect to their average is 1.45 against 1.40 in logarithmic unit of density and 303 against 291 \kms\  for RZA-radial and RZA3D respectively. Values are in agreement with results found previously in section 4.2 for mock constrained simulations. RZA3D constraints are stronger than RZA-radial ones. %\textcolor{blue}{This is best seen on Figure \ref{variance}. On that Figure, the cell-by-cell difference between the ratios, density over the average density, of the constrained simulations and of the random realizations are shown in function of the distance to the center of the box. More precisely, each random field obtained at z=0 is subtracted to the corresponding (same seed) constrained field at redshift null. Then the average of the differences over the 10 simulations is computed. Clearly, RZA-radial is less efficient at constraining inside the 30 \hMpc\ radius sphere. On the sphere where Virgo is supposed to be (10-15 \hMpc), there is a clear residual between the constrained and the random component only with RZA3D. With RZA-radial, the difference is flat, indicating that the random field dominates. If the Great Attractor region can be seen with both methods (peak at roughly 25 \hMpc), it has a higher density than Centaurus only with RZA3D. Beyond the edge of the cosmicflows-1 catalog, both types of constraints have an impact on the density field. However, because this first catalog is very shallow, we cannot pretend to study the cosmic variance in details beyond the 30 \hMpc\ boundary. At the edge of the box, the residual decreases and becomes quasi null for both methods. The random component is the only one left at the edge of the box. As a result, increasing the boxsize would not affect the constrained inner box. For instance, Virgo would still not be simulated properly with RZA-radial.}

\subsection{Constrained Simulations and Local Cosmography}
\label{subset:CS-CF1-WMAP7}

Since no Virgo cluster is simulated out of 10 RZA-radial simulations constrained by the observational peculiar velocity catalog,  the rest of the paper focuses only on the analysis of RZA3D constrained simulations. This section presents how compatible they are with the observed local cosmography.

\begin{table*} 
\begin{center}
\begin{tabular}{lllll}
\hline
 Case & Mass &  Average supergalactic  & Average distance to the & Nb of\\
 & &X,Y,Z position & observed Virgo cluster&  occurences\\
\hline
Observed & 4* &  -2.74, 12.0, -0.518  &  & \\
Virgo cluster & &&&\\
\hline
\fbox{RZA3D} Virgo (WMAP7) & 0.7 ; $\sigma=0.3$ & 1.23, 13.8, 2.2 ; $\sigma=1.2$  &  5.4;  $\sigma=1.2$  & \fbox{8/10}\\
 \fbox{RZA3D} Virgo (WMAP3)& 0.5 ; $\sigma=0.2$ & 1.30, 13.8, 0.70 ; $\sigma=1.0$  &  4.7;  $\sigma=0.8$  & \fbox{8/10}\\
\hline
\underline{RZA-radial}  Virgo (WMAP7) &  &  &    & \underline{0/10} \\
\underline{RZA-radial}  Virgo (WMAP3) &  &  &    & \underline{0/10} \\
%\hline
%\textbf{ Centaurus} & \textbf{6-7*}  &  \textbf{-30.3}, \textbf{13.3}, \textbf{-6.66} & &\\
%\hline
%RZA3D & 6 & -21.1, 12.8, -7.99 & 9.3 & 10 \\
    %          & $\sigma=3$  & &$\sigma=1.6$ &\\
\hline
\end{tabular}
\end{center}
\vspace{0.1cm}
\caption{Average parameters and standard deviations $\sigma$ for the halos representative of Virgo. (1) simulations in which the halos are looked for, 2) mass in h$^{-1}$ $10^{-14} \times$ solar mass. $^*$Estimation of the total (baryonic + dark matter) mass. (3) Average supergalactic coordinates X, Y and Z in \hMpc\  and standard deviation. (4) distance in \hMpc\ from the simulation halo to the observed Virgo location and standard deviation $\sigma$, (5) number of occurences in ten different simulations (if a halo similar to Virgo was found in a 6\hMpc\ sphere).}
\label{table1}
\end{table*}

\begin{figure*}
\centering
\vspace{-0.5cm}
\includegraphics[width=1\textwidth]{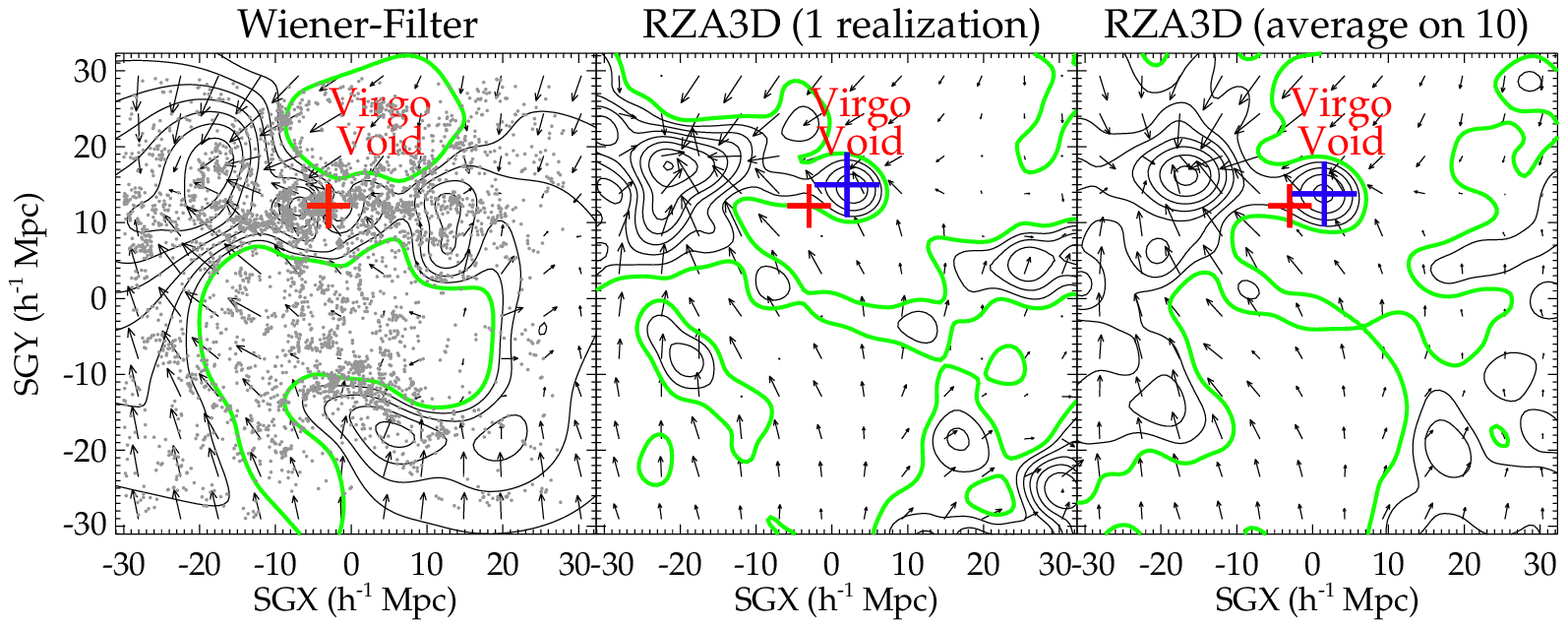}\\
\vspace{0.5cm}
\includegraphics[width=1\textwidth]{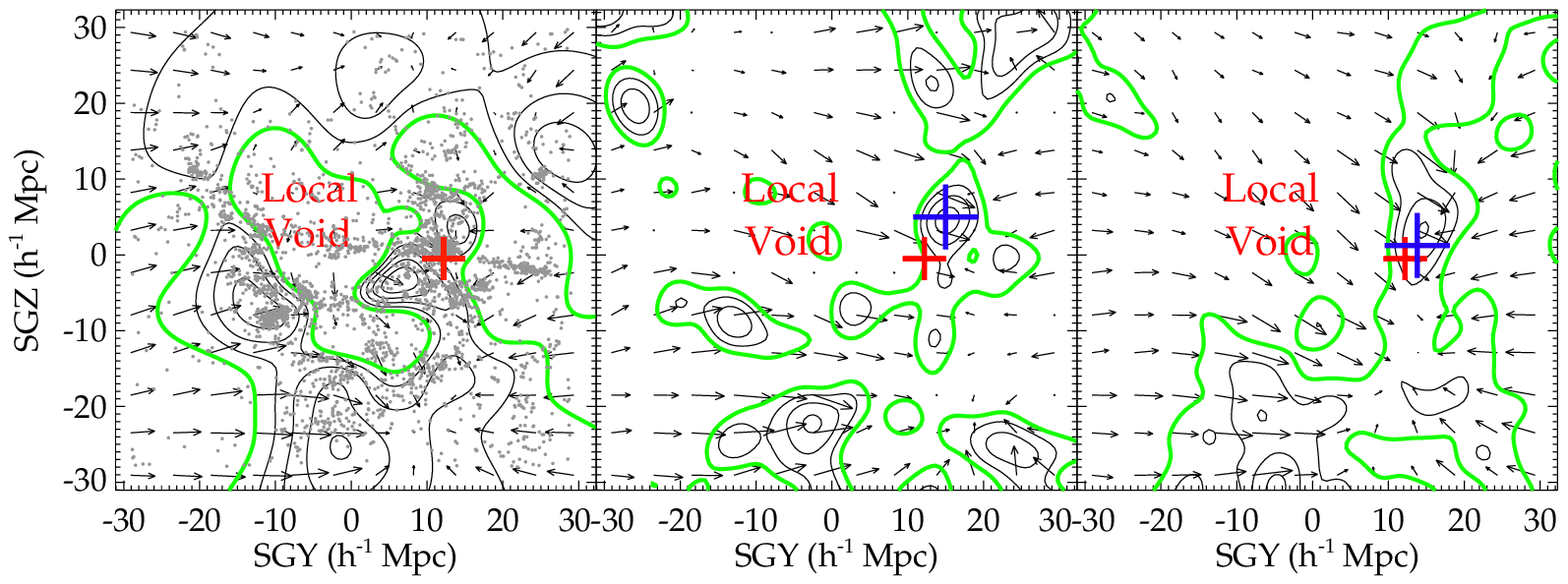}\\
%\vspace{0.5cm}
%\includegraphics[width=1\textwidth]{cf1_256_160cf1_RZA3D_256_160_s4cf1_RZA3D_256_160_stackedSGXZ00.eps}
\vspace{0.5cm}
\caption{XY, YZ supergalactic slices of the WF reconstruction (left), of one constrained simulation (middle) and of the average of 10 constrained simulations of the Local Universe within a 30 \hMpc\ radius sphere. The supergalactic slices are located at X = -2.5 and Z = 0 \hMpc\ to fit  the location of the Virgo cluster. The overdensity at 2 \hMpc\ Gaussian smoothing is represented with black iso-contours. The green contour stands for the mean density. The flows are shown with black arrows. In the XY supergalactic plane, the Virgo cluster and the Virgo Void are both reconstructed (left column) and simulated (middle column). Virgo is also visible next to the Local Void in the YZ supergalactic slice. In general Virgo is well simulated, at a similar location, whatever random component is used (right column). V8k (catalog of redshifts) galaxies are shown for reference as grey dots in a $\pm$ 10 \hMpc\ thick slice on the WF reconstruction. The red crosses locate Virgo in cosmicflows-1. The bigger blue crosses represents the (average) location of the Virgo-like halos.}
\label{wfsimucf1}
\end{figure*}

Figure \ref{wfsimucf1} displays the Wiener-Filter reconstruction of the cosmicflows-1 catalog (left), one single RZA3D simulation (middle column) and an average of ten RZA3D constrained simulations (right). Overdensity fields are smoothed by a Gaussian kernel of 2 \hMpc. Red crosses mark the Virgo cluster's position in the observed Universe (see Table \ref{table1} for the exact position). The Local Void and Virgo Void are also indicated. % The 20 most massive AHF halos of each simulation have been identified. The black dots represent those that are located within a $\pm 5 \hmpc$ slab centered on the plotted principal planes. 
The WF maps of cosmicflows-1 are taken as proxies to the actual Universe. They serve as targets for the constrained simulations, with the caveat that the WF provides only the linear overdensity field. The inner $R=30 \, \hmpc$ volume is dominated by the Local Supercluster. The general structure of the Local Universe including positions of the voids are quite well reproduced by a constrained simulation. The average over ten different realizations shows that in general the Virgo cluster region is well simulated at a similar location whatever random component is used. To quantify the reliability of the RZA3D technique in simulating the Virgo cluster and the area surrounding it, 1) the high density peak of the WF reconstruction is identified in each RZA3D simulation and 2) the Amiga halo finder is used to identify replicas of the cluster in the constrained simulations. We use the same process as with the mock catalog. There is a density peak at a similar location to the WF peak in 10 out of 10 RZA3D constrained simulations. The typical misplacement with respect to the WF peak is about 8-9 \hMpc\ with a standard deviation about 2 \hMpc . Although the supergalactic Y and Z components are very similar in the WF ($\sim$ 13 and 1 \hMpc) and in the simulations ($\sim$ 13 $\pm$ 1 and 3 $\pm$ 2 \hMpc), the error in position is very high because in the supergalactic X direction the shift in position with respect to the cosmicflows-1 Virgo cluster is negative in the WF ($\sim$ -3 \hMpc) while it is positive in the simulations ($\sim$ 4 \hMpc). Still, there is absolutely no density peak in the RZA-radial constrained simulations even when looking in a $\sim$ 10 \hMpc\ sphere centered on the WF density peak. The blue crosses in Figure \ref{wfsimucf1} stand for the positions (average positions, in the right column) of the Virgo-like halos. Table \ref{table1} provides details about masses, positions, error in position and standard deviations. For completeness, the table presents the results obtained in both WMAP7 and WMAP3 frameworks. Differences are negligible. A Virgo-like halo is present in 8 out of 10 simulations. By comparison, with RZA-radial, no replica of Virgo in  6 \hMpc\ spheres centered on the observational position was found. A synthetic Local Universe with a Virgo cluster using \emph{only observational peculiar velocities} is produced for the first time thanks to the WF/RZA/CR technique described in this paper.

%%%%%%%%%%%%%%%%%%%%%%%%%%%%%%%%%%%%%%%%%%%%%%%%%%%%%%%%%%%%%
%%%%%%%%%%%%%%%%%%%%%%%%%%%%%%%%%%%%%%%%%%%%%%%%%%%%%%%%%%%%%
%CONCLUSION%%%%%%%%%%%%%%%%%%%%%%%%%%%%%%%%%%%%%%%%%%%%%%%%%%%%%
%%%%%%%%%%%%%%%%%%%%%%%%%%%%%%%%%%%%%%%%%%%%%%%%%%%%%%%%%%%%%
%%%%%%%%%%%%%%%%%%%%%%%%%%%%%%%%%%%%%%%%%%%%%%%%%%%%%%%%%%%%%

\section{Conclusion}

The main aim of this paper is to perform numerical cosmological simulations constrained for the first time solely by an observational catalog of peculiar velocities. To build Initial Conditions (ICs), the Wiener-Filter (WF), the Reverse Zel'dovich Approximation (RZA) and the Constrained Realization (CR) techniques are successively applied.  A refinement is also added to the methodology. 

Since the cosmicflows-1 peculiar velocity catalog extends only out to about 30 \hMpc\ (radius), derived constrained simulations are subject to considerable cosmic variance. To study this effect, mock catalogs have been drawn from a previous constrained simulation which looks like the Local Universe and an ensemble of ten Initial Conditions have been constructed. The mock catalogs have been designed to mimic the observational catalog by including distance measurement errors and a large continuous zone without data (Zone of Avoidance due to our Galaxy extinction). The RZA algorithm with its new feature (called RZA3D) has been tested against these mocks and resulting simulations have been compared with the ones obtained with the original RZA version (called RZA-radial). The enhanced precision and reliability of the RZA3D method is validated. The methodology has been subsequently applied to the actual cosmicflows-1 catalog. An ensemble of ten constrained simulations has been constructed and analyzed. 
 %The first CLUES simulations \citep{2010arXiv1005.2687G} was not accounting for the (Zeldovich) displacements of the data points within the original WF/CR algorithm used to set up the constrained ICs. A partial remedy to the problem was devised by using the RZA methodology suggested and tested by \citet[RZA,][]{2013MNRAS.430..902D,2013MNRAS.430..912D,2013MNRAS.430..888D}. The original RZA method has been refined here.The mismatch between the targeted clusters and the position of the corresponding dark matter halos has been reduced below 5 \hMpc . There are only small shifts of about 2 \hMpc\ between the V8k redshift catalog and the simulations in the Y and Z supergalactic components. Along the X supergalactic axis, the higher shift of the simulated Centaurus is due to the absence of knowledge on the dynamics caused by the large bulk flow towards Shapley because it lies outside of the cosmicflows-1 data zone.
%The paper is setting the stage for a production mode of constrained simulations of the Local Universe from catalogs of peculiar velocities exclusively. 

The methodology succeeds in performing robust constrained simulations using \emph{only observational peculiar velocities} as constraints. 
The cosmicflows-1 catalog is still too shallow to enable constrained simulations that can reproduce all the main attractors and voids of the local dynamics. The recently published cosmicflows-2 catalog \citep{2013arXiv1307.7213T} contains more than 8,000 galaxy distances (1,800 in cosmicflows-1) and extends out to about 150 \hMpc. The technique reported in this paper is to be applied to this larger dataset with the aim of providing Initial Conditions for a constrained simulation of the universe in a 640 \hMpc\ box that will more thoroughly mimic observed large scale structure.

\section*{Acknowledgements}
We acknowledge help from and discussions with Timur Doumler. We thank the anonymous referee whose comments have contributed to improve this paper a lot. JS received support from the ''Projet d'avenir Lyon-St Etienne'' and from the "R\'egion Rh\^one-Alpes" via the Explora'doc grant. JS and HC acknowledge support from the Lyon Institute of Origins under grant ANR-10-LABX-66. RBT acknowledges support from NASA though the Spitzer Science Center {\it Cosmicflows with Spitzer} award and through the Astrophysical Data Analysis Program {\it Cosmicflows with WISE} award. YH has been partially supported by the Israel Science Foundation (13/08). SG and YH have been partially supported by the Deutsche Forschungsgemeinschaft under the grant $\rm{GO}563/21-1$. The simulations have been performed at the Leibniz Rechenzentrum (LRZ) in Munich.

\section*{Appendix}
\renewcommand{\thesubsection}{\Alph{subsection}}

\subsection{Constrained Realization technique}
The Constrained Realization technique (CR) is the optimal minimal variance estimator given a dataset and an assumed prior power spectrum. Data dominate the reconstruction in region where they are dense and accurate. On the opposite when they are noisy and sparse, the reconstruction is a prediction based on the assumed prior model. It schematically multiplies data by $\frac{Power Spectrum}{Power Spectrum + Error^2}$. Consequently, it would tend to the null field when data degrade if the fluctuations of an independent random realization are not added to compensate for the missing power spectrum. Without the random field, the result of the computation is the Wiener-Filter. Briefly, the overdensity $\delta^{CR}$ and velocity $\vec v^{CR}$ fields of constrained realizations are expressed in terms of the random realization fields $\delta^{RR}$, $\vec v^{RR}$ and the correlation matrixes. For a list of M constraints $c_i$:

\begin{equation}
\delta^{CR}(\vec r)=\delta^{RR}(\vec r)+\sum_{i=1}^M \langle\delta(\vec r)c_i\rangle\eta_i 
\label{eq1}
\end{equation}

\begin{equation}
 v_{\alpha}^{CR}=v_{\alpha}^{RR}(\vec r) + \sum_{i=1}^M \langle v_{\alpha}(\vec r)c_i\rangle \eta_i \quad with \quad \alpha=x,y,z
 \label{eq2}
 \end{equation}

where $\eta_i=\sum_{j=1}^M\langle C_i C_j\rangle^{-1}(C_j-\overline{C_j})$ are the components of the correlation vector $\eta$. $\overline C_i$ are random constraints with the noise and $C_i=c_i+\epsilon_i$ are mock or observational constraints plus their uncertainties. Hence, $\langle C_i C_j\rangle$ is equal to $\langle c_i c_j\rangle+\epsilon_i^2\delta_{ij}$ assuming errors statistically independent. The constraints can be either densities or velocities. The project Cosmicflows uses only peculiar velocities obtained from direct distance measurements. $\langle AB \rangle$ notations stand for the correlation functions involving the assumed prior power spectrum.\\
 
The associated correlation functions are given by: 
\[ \langle \delta(\vec r\, ') v_{\alpha} (\vec r \,'+\vec r) \rangle  = \frac{\dot a f}{(2 \Pi)^3}\int_0^\infty \frac{ik_{\alpha}}{k^2}P(\vec k) e^{-i\vec k.\vec r}d\vec k \]
\begin{equation} \; = -\dot a f r_{\alpha} \zeta (r) \end{equation}

\[  \langle v_{\alpha}(\vec r \,')v_{\beta}(\vec r\, '+\vec r)\rangle \;\; = \frac{(\dot a f)^2}{(2\Pi)^3}\int_0^\infty \frac{k_{\alpha}k_{\beta}}{k^4}P(\vec k) e^{-i \vec k .\vec r} d\vec k  \]
\begin{equation} = (\dot a f)^2 \Psi_{\alpha\beta} \end{equation}

where P is the assumed prior power spectrum.\\

Because data sample a typical realization of the prior model, i.e. the power spectrum, $\frac{\chi^2}{d.o.f}$ should be close to 1 where $\chi^2=\sum_{i=1}^M\sum_{j=1}^M C_i\langle C_i C_j \rangle^{-1} C_j$ and d.o.f is the degree of freedom. However, data include non-linearities which are not taken into account in the model. Consequently, a  $\sigma_{NL}$ such that $\langle C_i C_j \rangle = \langle c_i c_j \rangle+\delta_{ij}^k\epsilon_j^2+\delta_{ij}^k\sigma_{NL}^2$ is required to compensate for the non-linearities to drive $\frac{\chi^2}{d.o.f}$ closer to 1.\\

\subsection{The Reverse Zeldovich Approximation}

In the framework of Lagrangian perturbation theory, the comoving (Eulerian) position $\vec r$ of a data point can be written:

\begin{equation}
\vec r (t)= \vec q(\vec r) + \vec \Psi (\vec r , t)
\label{eqbefore}
\end{equation}

where $\vec q$ is the initial (Lagrangian) position and $\vec \Psi$ is the displacement field.\\

The Zel'dovich approximation consists in writing $\vec \Psi$ as the product of two independent functions of position ($\vec \Psi_0(\vec r)$) and time ($\vec D_+(t)$)  respectively.
\begin{equation}
\vec r(t) = \vec q(\vec r) + \vec D_+(t) \vec \Psi_0(\vec r)
\label{eqzel}
\end{equation}
where $ \vec D_+(t)$ is the growing mode of structures (the decaying mode is assumed to have reached zero).

The comoving (peculiar) velocity $\vec v = a \frac{d \vec r}{dt}$ (with $a$ the scale factor) is obtained deriving equation \ref{eqzel}:
\begin{equation}
\vec v = \dot a f \vec \Psi
\end{equation} 
where $\dot a$ is the time derivative of the scale factor and f is the growth rate $f=\frac{d (lnD_+)}{d (lna)}$.

At z=0, $a=1$ thus $\dot a$ = $H_0$. It results that nowadays,
\begin{equation}
\vec v = H_0 f \vec \Psi
\label{eqvpsi}
\end{equation}

Assuming that the original positions of protohalos $\vec r_{\,init}^{\,RZA}$ is approximately equal to $\vec q$, the equation to shift back galaxies at $\vec r$ at z=0 to their precursors' positions at $\vec r_{\,init}^{\,RZA}$ at higher redshifts is from equations \ref{eqbefore} and \ref{eqvpsi}:

\begin{equation}
\vec r_{\,init}^{\,RZA}=\vec r -\frac{\vec v}{H_0 f}
\end{equation}

%\subsection{Divergent Velocity Field}

%\begin{equation}
 %\vec \nabla_{\vec r} . \vec v (\vec r) = - \dot a f \delta (\vec r)
%\end{equation}

%%%%%%%%%%%%%%%%%%%%%%%%%%%%%%%%%%%%%%%%%%%%%%%%%%%%%%%%%%%%%%%%%%%%%%%%%%%%%%%%%%%%%%%%%%%%%%%%%%%%%%%%%%%%%%%%%%%%%%%%%%%%%%%%
%%%%%%%%%%%%%%%%%%%%%%%%%%%%%%%%%%%%%%%%%%%%%%%%%%%%%%%%%%%%%%%%
\clearpage

\bibliographystyle{mnras}

\bibliography{biblicomplete}

\end{document}